# Vertical Injection and Wideband Grating Coupler Based on Asymmetric Grating Trenches

Md Asaduzzaman[1,*], Robert J. Chapman[1,] Brett C. Johnson[1], and Alberto Peruzzo[1]

[1]Quantum Photonics Laboratory and Centre for Quantum Computation and Communication Technology, School of Engineering, RMIT University, Melbourne, Victoria 3000, Australia

*Correspondence: md.asaduzzaman@rmit.edu.au

A Silicon-on-insulator (SOI) perfectly vertical fibre-to-chip grating coupler is proposed and designed based on engineered subwavelength structures. The high directionality of the coupler is achieved by implementing step gratings to realize asymmetric diffraction and by applying effective index variation with auxiliary ultra-subwavelength gratings. The proposed structure is numerically analysed by using two-dimensional Finite Difference Time Domain (2D FDTD) method and achieves 76% (-1.19 dB) coupling efficiency and 39 nm 1-dB bandwidth.

Recently, the miniaturization of optoelectronic and photonic devices to a photonic integrated circuit (PIC) has gained significant momentum due to improved performance, enhanced reliability, lower energy consumption and lower manufacturing cost [1]. The concurrent advancement of silicon-on-insulator (SOI) that benefits from high contrast between the indices of silicon (Si) and its oxide has fuelled this progress further [2]. High index contrast allows dense integration of optical devices in single silicon (Si) substrate with multiple functionalities and interconnecting them using subwavelength-scale photonic waveguides [3]. One of the key challenges with such ultra-thin waveguides is to find a suitable method to interface them with external optical fibres. This is due to the large mismatch in shape and size of the fundamental modes of the waveguide and fibre. Such mismatch results in inefficient coupling of light between waveguide and fibre, demanding the development of efficient and wideband coupling mechanisms [4, 5].

Among various coupling mechanisms reported thus far, grating couplers (GC) are considered as the preferred choice due to its high compatibility with standard CMOS fabrication processes. Unlike facet coupling, GCs avoid post-fabrication complexities of dicing and polishing [6-8] and offers the flexibility to place it at arbitrary positions through the chip for wafer level testing and mass production [9, 10]. One important aspect of GCs is its alignment with interfacing optical fibre for effective light transporting. Although coupling efficiency (CE) and bandwidth (BW) in GCs has been widely studied in the recent past [11-14], most of these investigations relied on tilted positioning of the fibre causing increased complexity during the packaging process of the PICs. To overcome this problem, GCs with vertical fibre injection were proposed [15-19] with reported CEs < 70%. These CEs are achieved using various complex structures such as top/bottom reflectors, apodized, slanted, chirped gratings which often involved customized wafer and fabrication processes.

In this work, we propose a new GC structure for perfectly vertical coupling by creating asymmetric grating trenches with a combination of step and engineered subwavelength gratings that manipulate the effective refractive index to achieve higher directionality. The high CE and BW in this CMOS compatible structure can be achieved without the need of top or bottom reflector using commercially available standard SOI wafer.

The work reported in this article is organized as follows: Section I describes the design and working principles of the grating coupler with primary and secondary gratings optimized for vertical matching with single mode fibre (SMF). Section II describes the results and analysis of the proposed subwavelength engineered grating coupler. Section III presents optimization of the secondary gratings for maximizing the performance of the grating coupler under discussion. The performance optimization section is finally followed by concluding remarks in Section IV.

## I. Proposed Design Layout

Our design process starts with defining the grating parameters on a SOI wafer consisting of 1500 nm $SiO_2$ as buried oxide (BOX) layer and 220 nm Si as top layer that ensure single mode operations. The thickness of the BOX ($t_{BOX}$) is optimized for constructive interference between propagated and reflected waves [20]. The geometry of the proposed structure is shown in Fig. 1. The structure can be fabricated on standard SOI wafer using Focused Ion Beam (FIB) or few steps of etching with Electron Beam Lithography (EBL) process.

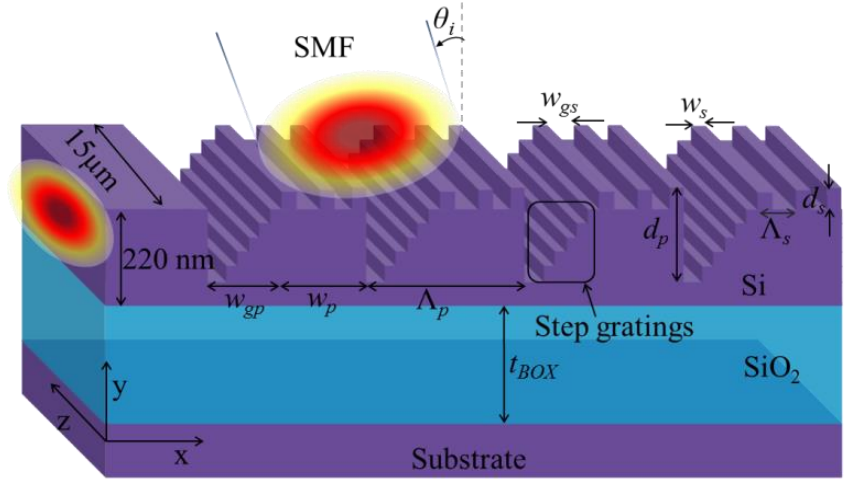

Fig. 1. Geometry of the proposed engineered subwavelength grating coupler (drawing not to scale).

The grating structure is categorised in two steps namely primary, denoted with subscript "*p*", and secondary, denoted with subscript "*s*". In order to obtain diffraction of light in gratings, we first require determining the period of primary grating ($\Lambda_p$) using equation (1)

$$\Lambda_p = \frac{\lambda}{n_{eff} - n_i \sin \theta_i} \qquad (1)$$

where $\lambda$ is the wavelength of the incident wave, $n_{eff}$ is the effective refractive index of structure, $n_i$ is the refractive index of the top cladding, and $\theta_i$ is the incident angle. A Gaussian shape TE polarized optical source with mode field diameter (MFD) of 10 μm, similar to that of standard SMF, is applied on the surface of the grating. Initially the $n_{eff}$ is calculated without the secondary gratings using fully vectorial mode solver and found to be 2.214 for the fundamental mode. By setting the targeted wavelength at λ=1550 nm, the primary grating period for the vertically incident is calculated as $\Lambda_p$ = 700 nm. Initially the duty cycle ($DC_p$) and etch depth ($d_p$) are set to 50% of the total grating period and top Si waveguide thickness respectively, which were found optimum later at $DC_p$ = 0.56, $d_p$ = 90 nm by using parameter sweep algorithm. The structure was then updated with the optimum values to find grating width, $w_p = (\Lambda_p * DC_p) = 392$ nm, and groove width, $w_{gp} = (\Lambda_p - w_p) = 308$ nm.

The number of diffraction orders that can exist for the grating period of 700 nm is calculated using the equation as follows:

$$n_{eff} \sin \theta_m = n_i \sin \theta_i - m \frac{\lambda}{\Lambda_p} \quad (2)$$

where $m$ is the diffraction order ($m = \pm 1, \pm 2 \ldots$), $\theta_m$ is the angle of diffraction of the $m^{th}$ order in material with refractive index of $n_{eff}$. It is found that only $m = -1$ diffraction order exists and diffracted at the angle of $\theta_{-1} \sim 90^0$ confirming that the $-1^{st}$ diffraction order is propagating along the waveguide. However, a portion of the diffracted power leaks through BOX and coupled into substrate. There also exist fundamental $0^{th}$ order which carries portion of incident power and propagates at different angle of $\theta_0 \sim -0.0939^0$ which implies that the $0^{th}$ order propagates almost vertically towards the substrate. Furthermore, there is substantial reflection at the grating surface for normal incident causing poor coupling efficiency. By appropriate adjustment of the grating structure the power with $0^{th}$ order can be coupled into -$1^{st}$ order [21-23] and also the coupling loss can be minimized by means of directionality enhancement to improve the coupling efficiency of the GC.

To increase the propagation towards the waveguide, the grating structure is remodelled with auxiliary ultra-subwavelength gratings as shown in Fig.1. The additional grating structure is named as the secondary grating, parameters of which is denoted as subscript with "$s$" (shown in Fig.1 and Fig.2). The design of the secondary grating is started with parameterization of the step gratings at one end of the groove of the primary gratings as shown in Fig.2. We use the slope of the primary groove to determine the thickness and width such that the top edge of each step becomes tangential with the slope so that the slope angle with each step grating equals to slope angle of the primary top groove that is $\varphi_p = \varphi_{s1} = .. = \varphi_{s5} = \tan^{-1}(w_{gp}/d_p)$. There are various combinations of the widths and depths of the step gratings depending on the number of gratings into the primary groove that fulfils the tangential conditions with the slope. In this design of primary groove depth and width of 90 nm and 308 nm respectively, we calculated the CE for various number of step gratings as shown in Fig 3. From the results it can be noted that the CE increases almost linearly with the number of steps in the primary groove until it nearly saturates after 4 steps. Similar trends also observed for the BW except slight downwards of the curve with increased number of step gratings.

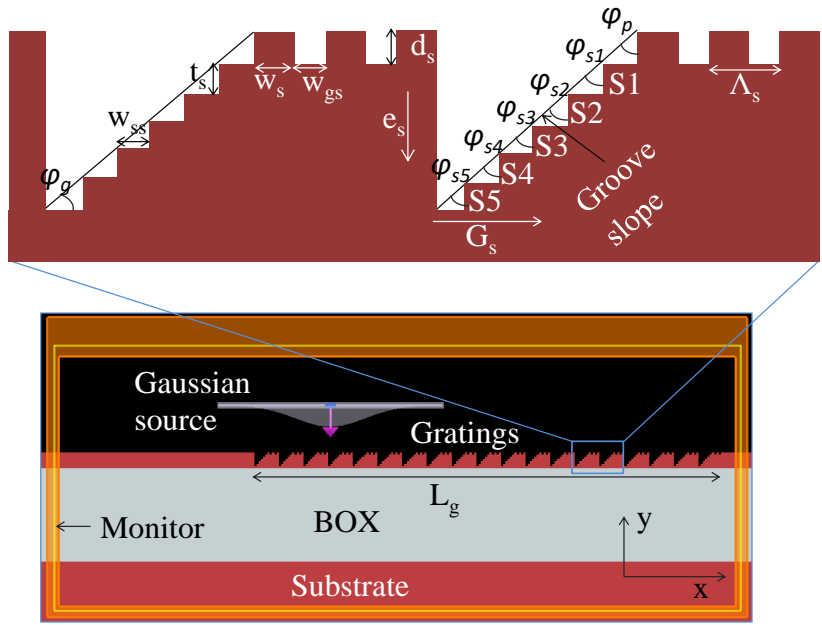

Fig. 2. 2D FDTD simulation environment setup. Inset: geometry of the secondary gratings (drawing not to scale).

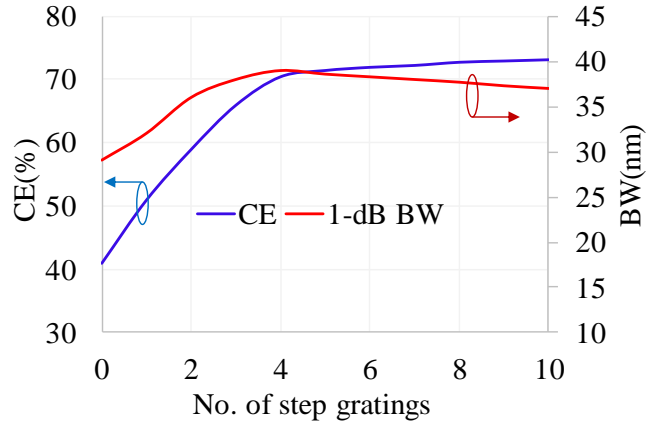

Fig. 3. Coupling efficiency at wavelength of 1550 nm and bandwidth for number of steps gratings into the primary grooves.

Such features can be attributed to the fact that the insertion of additional steps into the primary groove modified the effective refractive index which causes lower CE for longer wavelengths. Hence a minor downward trend of the BW curve. Therefore, we choose the number of step gratings $N = 4$ to fulfil the tangential condition and the thickness ($t_s$) and width ($w_{ss}$) are calculated as $t_s = [d_p / (N+1)] = 18$ nm and $w_{ss} = (w_{gp} / N) = 77$ nm.

It can be noted that this is the maximum width for a step grating for constant thickness and we rename it $w_{ss\text{-}max}$. In such a case the groove space ($G_s$) equal to $w_{ss}$. The relationships of $\varphi_p$, $G_s$ and $w_{ss}$ can be expressed by the following equations:

$$w_{ss} = \frac{\tan(\varphi_p) \cdot d_p}{N} \tag{3}$$

$$G_s = w_{gp} - (w_{ss} \cdot N - w_{ss-\max}) \tag{4}$$

Similarly, the relationships of the primary bottom groove angle ($\varphi_g$), etch space ($e_s$) and $t_s$ can be expressed by the following equations:

$$t_s = \frac{\tan(\varphi_g) \cdot w_{gp}}{N+1} \tag{5}$$

$$e_s = d_p - [t_s \cdot (N+1)] \tag{6}$$

The step grating width and corresponding groove space obtained using equations (3) and (4) for various primary top groove angles are shown in Fig.4 (a) while Fig.4 (b) shows the thickness of step gratings and corresponding etch space calculated using equations (5) and (6) for various primary bottom groove angles. It can be noted that when $\varphi_p = 0$, $G_s = w_{gp} = 308$ nm and when $\varphi_g = 0$, $e_s = d_p = 90$ nm. In such case the structure refers to the primary grating design.

The design of top secondary grating in the ridge of primary grating is started with defining the grating width ($w_s$) and etch depth ($d_s$) similar to that of step grating which are $w_s = 77$ nm and $d_s = 18$ nm. By setting the duty cycle ($DC_s$) at 50%, the secondary grating period is calculated as $\Lambda_s = 154$ nm with secondary groove width $w_{gs} = 77$ nm as well. The asymmetric trenches are developed by introducing the step gratings at the one end of the primary groove while another end is kept unaltered as shown in inset of Fig.2. The secondary grating period is so small ($\Lambda_s \ll \lambda$) that it acts as a homogeneous medium to the incident waves. Therefore, such gratings do not cause any additional diffraction order. The secondary grating manipulates the effective refractive index for propagation directionality. The effect of the structural parameters of the secondary grating including groove space on the coupling efficiency and bandwidth is discussed in section III.

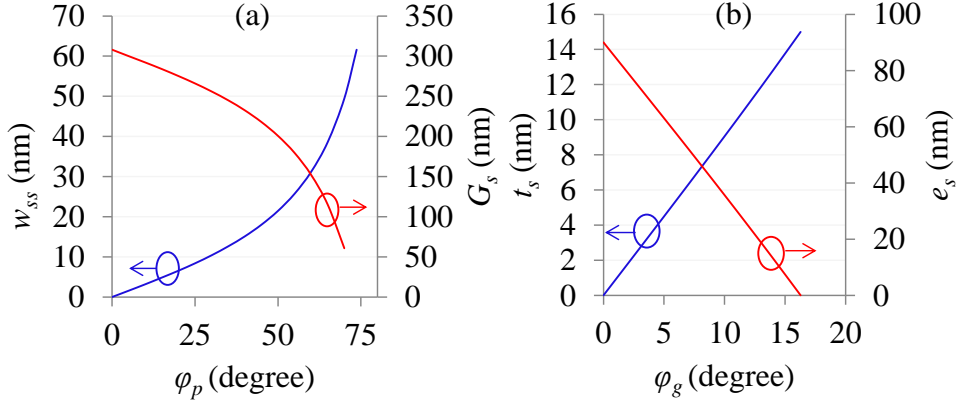

Fig. 4. (a) The relationships of $\varphi_p$, $G_s$ and $w_{ss}$ and (b) the relationships of $\varphi_g$, $e_s$ and $t_s$.

## II: Results and Discussion

The structure is simulated based on two-dimensional Finite Difference Time Domain (2D FDTD) method using commercially available software package Lumerical FDTD Solutions. Although grating couplers are 3 dimensional, the 3D structure is approximated using 2D simulation with high accuracy as the width of the grating waveguide (15 μm) is much larger compare with the height of the waveguides (220 nm) and wavelength (1550 nm) [24]. The grid spacing is set as small as dx = dy = 5 nm for high accuracy of the results. Perfectly Matched Layer (PML) is used to prevent reflections from the boundary by implementing impedance matching between simulation region and its materials [25]. The 2D FDTD simulation environment setup is shown in Fig.2.

The structure is first simulated with only primary gratings and the E-field distribution along the grating structure and CE at wavelength of 1550 nm obtained from 2D FDTD simulation are shown in Fig. 5. Fig.5 (b) shows that around 43% (-3.665 dB) of incident light could couple into nano waveguide. The loss can be attributed to the fact that a large amount of leakage light coupled into substrate through BOX and there exists backward scattering as shown in Fig.5 (a) which causes poor CE of the GC. Fig.5 (c) shows the CE over a wavelength range of 1500 - 1600 nm with centre at 1550 nm where the peak of the efficiency occurred and 1-dB coupling bandwidth of 29 nm is achieved. The backward scattering happens due to the symmetrical structure of the gratings. The initial symmetry of the gratings is broken by placing the optical source toward the waveguide other than at midpoint of grating length ($L_g$) as shown in Fig.2. Therefore, higher forward propagation happens compared to backward propagation. However, there is still considerable amount of light at backward direction and that can be minimized by further breaking the symmetry for perfectly vertical incident with engineered ultra-subwavelength grating structure.

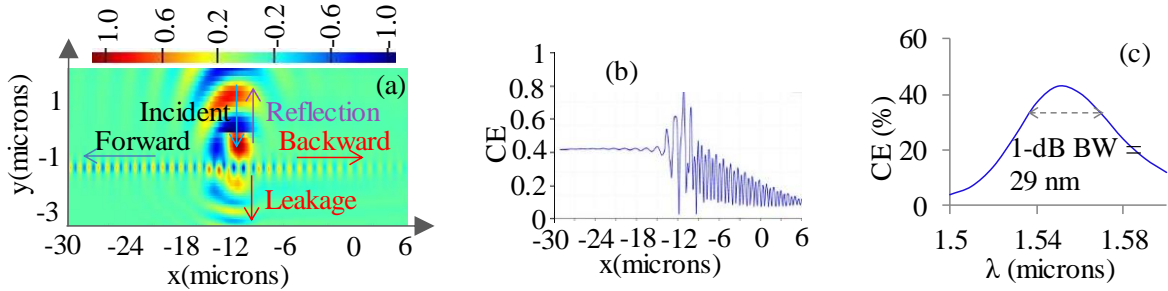

Fig. 5. (a) E-field distribution along grating structure (b) coupling efficiency at wavelength of 1550 nm and (c) coupling bandwidth.

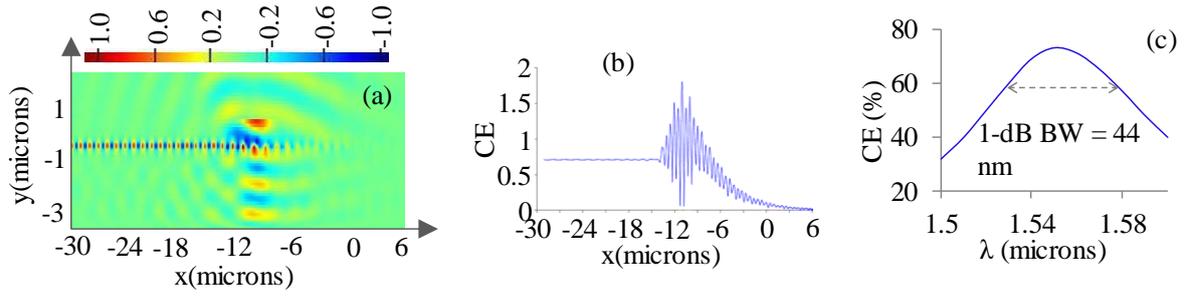

Fig. 6. (a) E-field distribution along grating structure (b) coupling efficiency at wavelength of 1550 nm and (c) coupling bandwidth for grating coupler with secondary ultra-subwavelength grating.

The proposed GC with secondary gratings including step gratings was then once again simulated by 2D FDTD with the same simulation environment setup used as before. The E-field distribution and CE obtained are shown in Fig.6. The light that coupled with substrate through BOX is reduced in the redesigned structure as can be seen comparing E-field distributions in Fig.5 (a) and Fig.6 (a). The directionality of the power propagation in subwavelength grating coupler is improved significantly. With such engineered grating structure, the CE of ~70% is achieved as shown in Fig.6 (b). Coupling bandwidth is shown in Fig.6 (c) which shows that peak coupling efficiency occurred at 1550 nm with 1-dB bandwidth of ~ 44 nm.

Fig.7 shows the reflection and transmission characteristics of the coupler in terms of scattering parameters (S-parameters). The reflection at the incident is always lower than 5% for wavelengths of 1500–1600 nm in GC with our proposed engineered ultra-subwavelength grating as shown in Fig.7 (b). It also can be noted that only 2% reflection happens at 1550 nm wavelength for which the structure is designed. Whereas reflections vary between 18%-30% for the same wavelength range in GC with only primary grating with peak at around 1520 nm wavelength as shown in Fig.7 (a).

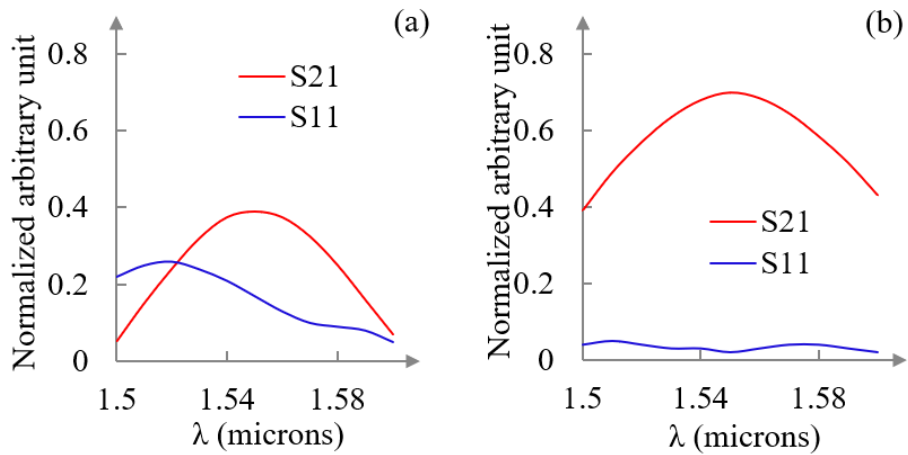

Fig. 7. Reflection (S11) and transmission (S21) with different wavelengths for vertical incident in (a) GC with only primary grating (b) GC with engineered sub-wavelength grating. Vertical axis is normalized to incident field intensity.

## III: Performance Optimization of the Secondary Gratings

In this section we investigate the performance of the coupler with various secondary grating parameters. First, the grating period ($\Lambda_s$) is varied between 100 nm and 140 nm for coupling efficiency over a range of wavelength as shown in Fig.8. The coupling spectra are almost similar for grating periods from 120 nm to 140 nm except the coupling peak moves slightly towards higher wavelength among those grating periods. However, the continuous decrement of the grating period results in the drop of CE as shown in Fig.8 that at grating period of 110 nm the CE falls and further dropping can be observed for grating period of 100 nm.

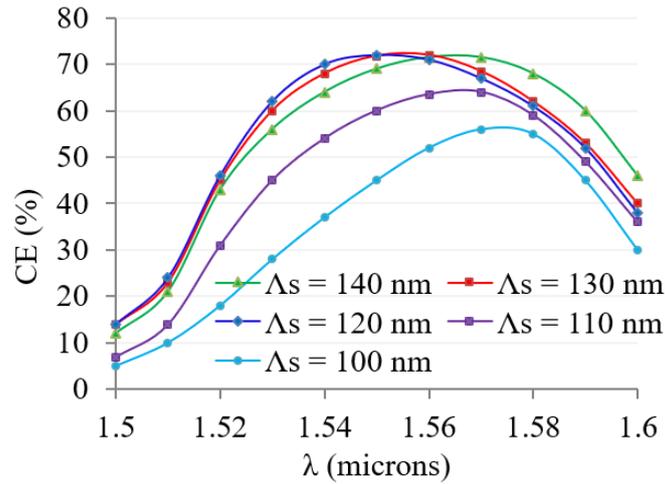

Fig. 8. Coupling efficiency vs wavelength for various secondary grating period.

The structure is simulated for different etch depths from 5 nm to 25 nm with 5 nm interval of secondary grating and the results are shown in Fig.9 (a). Results indicate that the CE with thinner etch depths exceeds the CE with deeper depths over the shorter wavelength region, whereas opposite happens with longer wavelengths. However, for etch depth of 10-20 nm the resonance occurs around 1550 nm wavelength at which the structure is designed. Most flat spectral responses occur at the etch depths of 10 nm, 15nm and 20 nm. Among them for $d_s = 18$ nm, the maximum CE happens with peak of ~ 72% at the wavelength of 1550 nm. The CE and bandwidth of the grating coupler for different duty cycle of the secondary grating is shown in Fig.9 (b). It is found that the maximum CE occurs at the duty cycle ($DC_s$) of 40%. The results also show that for lower duty cycle more light with shorter wavelength couples into waveguide whereas for higher duty cycle CE is high for light with longer wavelength.

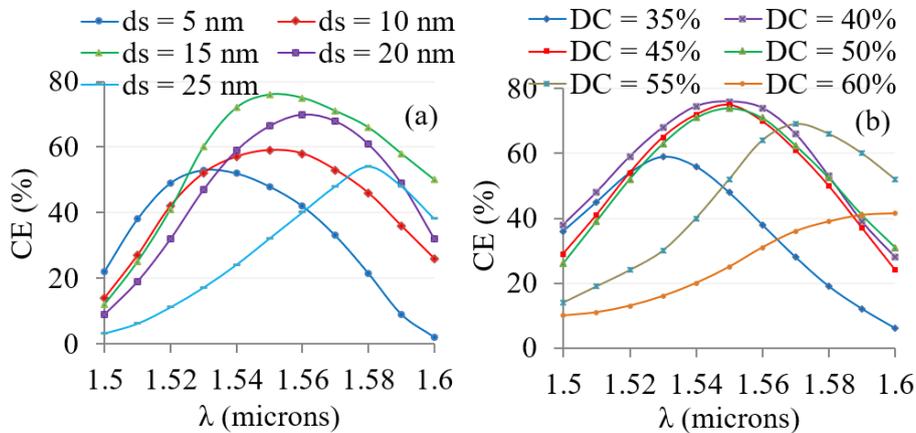

Fig. 9. Coupling efficiency vs wavelength for various (a) secondary etch depth ($d_s$) and (b) secondary duty cycle ($DC_s$).

Based on the relationships obtained from equations (3)-(6), Fig.10 shows the CE at wavelength of 1550 nm for the dimension of step gratings. CE for the various widths with corresponding groove spaces is shown in Fig.10 (a) which shows that over 75% CE is achieved for widths of 30-50 nm with peak of 76% at 38.6 nm which corresponds to groove space of 115 nm. Fig.10 (b) shows the CE obtained for various thicknesses of the step gratings. For increasing $t_s$ in the range of 0-15 nm, CE increases linearly while $e_s$ decreases which implies that there should not be any gap on the wall of groove above the first step grating S1 (see Fig.2). The results also show that if $w_{ss}$ and $t_s$ break the condition for the slope tangential, then CE drops as can be seen in Fig.10 (a) and (b).

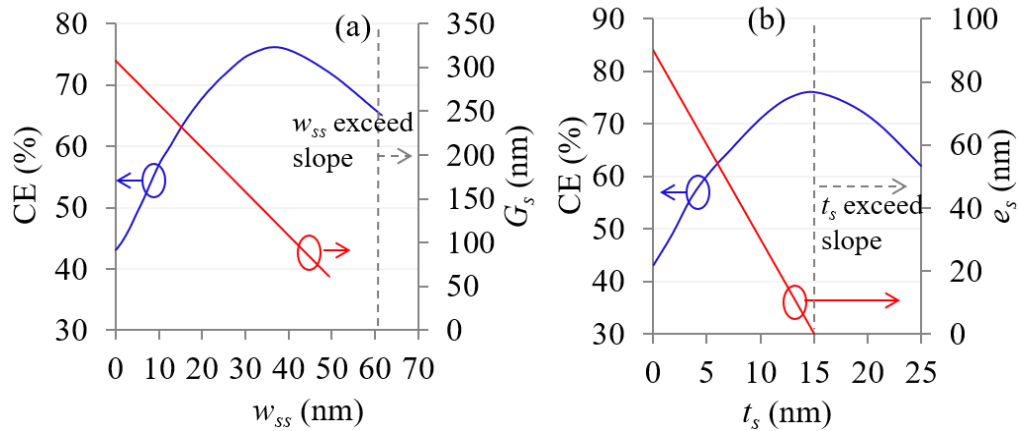

Fig. 10. Coupling efficiency for various (a) widths and corresponding groove spaces (b) thicknesses and corresponding etch spaces at 1550 nm wavelength.

Table 1: Optimized secondary grating parameters

| $\Lambda_s$ (nm) | $DC_s$ | $w_s$ (nm) | $w_{gs}$ (nm) | $d_s$ (nm) | $w_{ss}$ (nm) | $t_s$ (nm) |
|---|---|---|---|---|---|---|
| 154 | 0.4 | 62 | 92 | 15 | 38.6 | 15 |

Finally, the estimated coupling efficiency over the wavelength range of 1500-1600 nm for grating coupler with optimized secondary grating parameters as of in Table 1 is shown in Fig.11. The CE and BW increases with the number of step gratings in the primary groove up to 4 steps and nearly saturates with higher numbers as shown in Fig. 11. Significant increment of CE can be observed from step number 1 to 4 but are similar CE for steps 4 and 5. The peak coupling efficiency of ~ 76% is achieved at wavelength of 1550 nm with 1-dB coupling bandwidth of ~39 nm for 4 step gratings in the primary groove. Although the efficiency is improved with optimized structure, the bandwidth is little lower than that of obtained in Fig.6 (c). This is due to fact that the modified effective refractive index causes lower coupling efficiency for longer wavelengths. This once again proves that there always exists compromise between coupling efficiency and coupling bandwidth in a grating coupler.

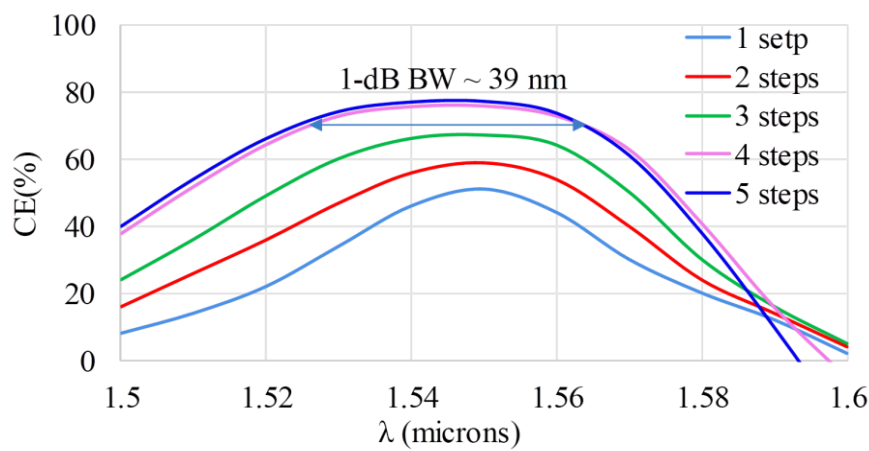

Fig. 11. Coupling efficiency vs. wavelength for optimized grating coupler with various numbers of step gratings in the primary groove.

## IV: Conclusion

In this work, a perfectly vertical grating coupler is proposed for light coupling between SOI waveguides and optical fibres. The symmetrical propagation nature of the diffracted waves from grating is broken by incorporating asymmetric grating trenches with step gratings. The directionality of the coupler is also boosted by implementing effective index variation of the primary grating achieved by combination of auxiliary ultra-subwavelength gratings and step gratings. In this design we have achieved coupling efficiency as high as 76% (-1.19 dB) and 1-dB coupling bandwidth of 39 nm. Such efficient and broadband perfectly vertical grating couplers are significantly advantageous in high dense photonic packaging.

**Reference**

1. M. Asghari and A. V. Krishnamoorthy, "Silicon Photonics: Energy-efficient communication", *Nat. Photonics* 5(5), 268 (2011).
2. R. Soref, "The past, present, and future of silicon photonics," IEEE J. Sel. Top. Quantum Electron. 12(6), 1678–1687 (2006).
3. Xiaochuan Xu, Harish Subbaraman, John Covey, David Kwong, and A. Hosseini, "Complementary metal-oxide-semiconductor compatible high efficiency subwavelength grating couplers for silicon integrated photonics," Appl. Phys. Lett. 101, 0311091, (2012).
4. S. Janz, P. Dalacu, D. Delˆage, A. Densmore, A. Lamontagne, B. Picard, M. Post, E. Schmid, J. Waldron, D. Xu, K.P. Yap, and W. Ye, "Microphotonic elements for integration on the silicon-on-insulator waveguide platform," IEEE J. Sel. Top. Quantum Electron. 12(6), 1402–1415 (2006).
5. G. Roelkens, D. Vermeulen, D. Van Thourhout, S. B. R. Baets, P. G. P. Lyan, and J.-M. Fédéli, "High efficiency diffractive grating couplers for interfacing a single mode optical fiber with a nanophotonic silicon-on-insulator waveguide circuit," Appl. Phys. Lett. 92, 1311101 (2008).
6. T. Shoji, T. Tsuchizawa, T. Watanabe, K. Yamada, and H. Morita, "Low loss mode size converter from 0.3 μm square Si wire waveguides to singlemode fibres," Electron. Lett. 38, 1669-1670 (2002).
7. V. R. Almeida, R. R. Panepucci, and M. Lipson, "Nanotaper for compact mode conversion," Opt. Lett. 28, 1302-1304 (2003).
8. S. McNab, N. Moll, and Y. Vlasov, "Ultra-low loss photonic integrated circuit with membrane-type photonic crystal waveguides," Opt. Express 11, 2927-2939 (2003).
9. Neil Na, Harel Frish, I-Wei Hsieh, Oshrit Harel, Roshan George, Assia Barkai, and Haisheng Rong, "Efficient broadband silicon-on-insulator grating coupler with low backreflection," Opt. Lett. 36, 2101-2103 (2011).
10. T. K. Saha and Weidong Zhou, "High efficiency diffractive grating coupler based on transferred silicon nanomembrane overlay on photonicwaveguide," J. Phys. D: Appl. Phys. 42, 085115, 1-9, (2009).
11. J. V. Galan, P. Sanchis, J. Blasco, J. Marti, "Study of High Efficiency Grating Couplers for Silicon-Based Horizontal Slot Waveguides", IEEE Photon. Technol. Lett. 20, 985-987 (2008).

12. X. Chen, K. Xu, Z. Cheng, C. K. Y. Fung, and H. K. Tsang, "Wideband subwavelength gratings for coupling between silicon-on-insulator waveguides and optical fibers," Opt. Lett. 37(17), 3483–3485 (2012).
13. X. Xu, H. Subbaraman, J. Covey, D. Kwong, A. Hosseini, and R. T. Chen, "Colorless grating couplers realized by interleaving dispersion engineered subwavelength structures," Opt. Lett. 38(18), 3588–3591 (2013).
14. Md. Asaduzzaman, Masuduzzaman Bakaul, Stan Skafidas, Md. Rezwanul Haque Khandokar, "Compact Silicon Photonic Grating Coupler With Dual-Taper Partial Overlay Spot-Size Converter", IEEE Photonics Journal, Vol. 9, No. 2, (2017).
15. Xia Chen, Chao Li, Hon Ki Tsang, "Fabrication-Tolerant Waveguide Chirped Grating Coupler for Coupling to a Perfectly Vertical Optical Fiber", IEEE Photon. Technol. Lett. 20, No.23, 1914-1916, (2008).
16. John Covey, Ray T. Chen, "Efficient perfectly vertical fiber-to-chip grating coupler for silicon horizontal multiple slot waveguides", Opt. Express 11, No.9, 10886-10896 (2013).
17. Siya Wang, Yue Hong, Yuntao Zhu, Jingye Chen, Shengqian Gao, Xinlun Cai, Yaocheng Shi, Liu Liu, "Compact high-efficiency perfectly-vertical grating coupler on silicon at O band", Opt. Express 25, No.18, 22032-22037 (2017).
18. G. Roelkens, D. V. Thourhout, R. Baets, "High efficiency grating coupler between silicon-on-insulator waveguides and perfectly vertical optical fibers", Opt. Lett. 32(11), 1495–1497 (2007).
19. Zan Zhang, Xiaotao Shan, Beiju Huang, Zanyun Zhang, Chuantong Cheng, Bing Bai, Tianxi Gao, Xiaobo Xu, Lin Zhang and Hongda Chen, "Efficiency Enhanced Grating Coupler for Perfectly Vertical Fiber-to-Chip Coupling", Materials 2020, 13, 2681; doi:10.3390/ma13122681.
20. Md. Asaduzzaman, Masuduzzaman Bakaul, Stan Skafidas, Md. Rezwanul Haque Khandokar, "Multiple layers of silicon–silica (Si–SiO2) pair onto silicon substrate towards highly efficient, wideband silicon photonic grating coupler", Opt. Quant. Electron. (2016) 48:478, DOI 10.1007/s11082-016-0746-0.
21. Chubing Peng and William A Challener, "Input-grating couplers for narrow Gaussian beam: influence of groove depth", Opt. Express 12, No. 26, 6481, (2004).
22. T. Clausnitzer, T. Kämpfe, E.-B. Kley and A. Tünnermann, "An intelligible explanation of highly-efficient diffraction in deep dielectric rectangular transmission gratings", Opt. Express 13, No. 26, 10448, (2005).
23. Martin Rumpel, Michael Moeller, Christian Moormann, Thomas Graf, and Marwan Abdou Ahmed, "Broadband pulse compression gratings with measured 99.7% diffraction efficiency", Opt. Express 39, No. 2, 323, (2014).
24. Dirk Taillaert, Peter Bienstman, and Roel Baets, "Compact efficient broadband grating coupler for silicon-on-insulator waveguides", Opt. Lett. 29, 2749 (2004).
25. J. P. Berenger, Perfectly Matched Layer (PML) for Computational Electromagnetics. Morgan & Claypool Publishers, (2007).

Author contributions: Md Asaduzzaman conceives the idea of the work and major calculations and analysis performed by him. Alberto Peruzzo has supervised the work while Robert J. Chapman and Brett C. Johnson assisted with structure designs and fabrications processes. All authors have contributed in discussion of the results and writing of the manuscript.

Competing financial interests:
The authors declare no competing financial interests.